\documentclass[aps,prl,onecolumn,amsmath,amssymb,superscriptaddress]{revtex4}
\usepackage{graphicx}
\usepackage{amssymb}
\usepackage{natbib}
\usepackage{color}

\newcommand{\beq}{\begin{eqnarray}}
\newcommand{\eeq}{\end{eqnarray}}
\newcommand{\ys}{\textcolor{black}}
\newcommand{\yy}{\textcolor{black}}

\begin{document}
\title{High-energy magnetic excitations from heavy quasiparticles in CeCu$_2$Si$_2$}
\affiliation{Department of Physics and Astronomy, Rice University, Houston, Texas 77005, USA}
\affiliation{MOE Key Laboratory of Materials Physics and Chemistry under Extraordinary Conditions and Shaanxi Key Laboratory of Condensed Matter Structures and Properties, School of Physical Science and Technology, Northwestern Polytechnical University, Xian 710072, China}
\affiliation{Department of Physics, University of California, Berkeley, California 94720, USA}
\affiliation{Materials Sciences Division, Lawrence Berkeley National Laboratory, Berkeley, California 94720, USA}
\affiliation{Center for Correlated Matter and Department of Physics, Zhejiang University, Hangzhou 310058, China}
\affiliation{National Research Council, Chalk River, Ontario K0J 1JO, Canada}
\affiliation{Beijing National Laboratory for Condensed Matter Physics and Institute of Physics, Chinese Academy of Sciences, Beijing 100190, China}
\affiliation{University of Chinese Academy of Sciences, Beijing 100049, China}
\affiliation{Leibniz-Institut f\"{u}r Festk\"{o}rper- und Werkstoffforschung (IFW) Dresden, Postfach 270116, D-01171 Dresden, Germany}
\affiliation{NIST Center for Neutron Research, National Institute of Standards and Technology, Gaithersburg, MD 20899, USA}
\affiliation{Songshan Lake Materials Laboratory, Dongguan, Guangdong 523808, China}

\author{Yu Song}
\email{yusong\_phys@zju.edu.cn}
\affiliation{Department of Physics and Astronomy, Rice University, Houston, Texas 77005, USA}
\affiliation{Department of Physics, University of California, Berkeley, California 94720, USA}
\affiliation{Materials Sciences Division, Lawrence Berkeley National Laboratory, Berkeley, California 94720, USA}
\affiliation{Center for Correlated Matter and Department of Physics, Zhejiang University, Hangzhou 310058, China}
\author{Weiyi Wang}
\affiliation{Department of Physics and Astronomy, Rice University, Houston, Texas 77005, USA}
\author{Chongde Cao}
\email{caocd@nwpu.edu.cn}
\affiliation{MOE Key Laboratory of Materials Physics and Chemistry under Extraordinary Conditions and Shaanxi Key Laboratory of Condensed Matter Structures and Properties, School of Physical Science and Technology, Northwestern Polytechnical University, Xian 710072, China}
\author{Zahra Yamani}
\affiliation{National Research Council, Chalk River, Ontario K0J 1JO, Canada}
\author{Yuanji Xu}
\affiliation{Beijing National Laboratory for Condensed Matter Physics and Institute of Physics, Chinese Academy of Sciences, Beijing 100190, China}
\affiliation{University of Chinese Academy of Sciences, Beijing 100049, China}
\author{Yutao Sheng} 
\affiliation{Beijing National Laboratory for Condensed Matter Physics and Institute of Physics, Chinese Academy of Sciences, Beijing 100190, China}
\affiliation{University of Chinese Academy of Sciences, Beijing 100049, China}
\author{Wolfgang L\"{o}ser}
\affiliation{Leibniz-Institut f\"{u}r Festk\"{o}rper- und Werkstoffforschung (IFW) Dresden, Postfach 270116, D-01171 Dresden, Germany}
\author{Yiming Qiu}
\affiliation{NIST Center for Neutron Research, National Institute of Standards and Technology, Gaithersburg, MD 20899, USA}
\author{Yi-feng Yang}
\affiliation{Beijing National Laboratory for Condensed Matter Physics and Institute of Physics, Chinese Academy of Sciences, Beijing 100190, China}
\affiliation{University of Chinese Academy of Sciences, Beijing 100049, China}
\affiliation{Songshan Lake Materials Laboratory, Dongguan, Guangdong 523808, China}
\author{Robert J. Birgeneau}
\affiliation{Department of Physics, University of California, Berkeley, California 94720, USA}
\affiliation{Materials Sciences Division, Lawrence Berkeley National Laboratory, Berkeley, California 94720, USA}
\author{Pengcheng Dai}
\email{pdai@rice.edu}
\affiliation{Department of Physics and Astronomy, Rice University, Houston, Texas 77005, USA}

\begin{abstract}	
Magnetic fluctuations is the leading candidate for pairing in cuprate, iron-based and heavy fermion superconductors. 
This view is challenged by the recent discovery of nodeless superconductivity in CeCu$_2$Si$_2$, and calls for a detailed understanding of the corresponding magnetic fluctuations. Here, we mapped out the magnetic excitations in \ys{superconducting (S-type)} CeCu$_2$Si$_2$ using inelastic neutron scattering, finding a strongly asymmetric dispersion for $E\lesssim1.5$~meV, which at higher energies evolve into broad columnar magnetic excitations that extend to $E\gtrsim 5$ meV. While low-energy magnetic excitations exhibit marked three-dimensional characteristics, the high-energy magnetic excitations in CeCu$_2$Si$_2$ are almost two-dimensional, reminiscent of paramagnons found in cuprate and iron-based superconductors. By comparing our experimental findings with calculations in the random-phase approximation,we find that the magnetic excitations in CeCu$_2$Si$_2$ arise from quasiparticles associated with its heavy electron band, which are also responsible for superconductivity. Our results provide a basis for understanding magnetism and superconductivity in CeCu$_2$Si$_2$, and
demonstrate the utility of neutron scattering in probing band renormalization in heavy fermion metals.   
\end{abstract}

\maketitle
\section{Introduction}
The discovery of superconductivity in CeCu$_2$Si$_2$ \cite{FSteglich} marked the beginning of decades-long intense research into unconventional superconductivity \cite{DJScalapino2012,GStewart2017}, encompassing cuprate \cite{PALee2006,BKeimer2015,Sidis}, iron-based \cite{GStewart2011,PDai2015,RMFernandes2014,QMSi2016} and heavy fermion superconductors \cite{CPfleiderer2009,OStockert2012,BDWhite2015}. The proximity of superconductivity to antiferromagnetic (AF) quantum critical points (QCP) in these systems implicate AF fluctuations that proliferate at the AF QCP as the pairing glue, leading to unconventional superconductivity with a sign-changing superconducting order parameter \cite{DJScalapino2012,TMoriya2003,GStewart2017}. 

Experimental evidence for magnetically driven superconductivity in these systems include: (i) reduction of the magnetic exchange energy is much larger than the superconducting condensation energy \cite{ EDemler1998,HWoo2006,CStock2008,OStockert2011,MWang2013,JLeiner2014,SCarr2016,TDahm2009}; (ii) the observation of a spin resonance mode, which in the spin-exciton scenario indicates a sign-changing superconducting order parameter \cite{DJScalapino2012,MEschrig2006,GYu2009,weber2019}; and (iii) the persistence of two-dimensional (2D) high-energy AF fluctuations that resemble spin waves in magnetically ordered parent compounds \cite{RJBirgeneau2006,OJLipscombe2007,MFujita2012,MLeTacon2011,MLeTacon2013,MPMDean2013,MLiu2012,KJZhou2013,MWang2013}. 

From an empirical perspective, it is important to identify whether these features are common in different unconventional superconductors, so that ingredients for a unified pairing mechanism may be established. 
Of the above observations, (i) is model-independent and has been verified for cuprate, iron-based and heavy fermion superconductors \cite{ EDemler1998,HWoo2006,CStock2008,OStockert2011,MWang2013,JLeiner2014,SCarr2016,TDahm2009}; (ii) the spin resonance mode has been found in cuprate, iron-based and heavy fermion superconductors, but their spin-excitonic nature needs to be separately tested and requires a quasiparticle origin of the magnetic excitations \cite{DJScalapino2012,MEschrig2006,GYu2009}; while (iii) \ys{has} been established for cuprate and iron-based superconductors \cite{RJBirgeneau2006,OJLipscombe2007,MFujita2012,MLeTacon2011,MLeTacon2013,MPMDean2013,MLiu2012,KJZhou2013,MWang2013}, magnetic excitations in heavy fermion superconductors such as CeCu$_2$Si$_2$ are strongly three-dimensional (3D) at low energies \cite{OStockert2011}, and it is unclear whether they become 2D at higher energies.

CeCu$_2$Si$_2$ (S-type) is an archetypal heavy fermion unconventional superconductor, and is naturally located near a 3D AF QCP \cite{PGegenwart1998,JArndt2011}. Upon the introduction of slight Cu deficiencies the system can be tuned to AF order (A-type), with an ordering vector $\tau\approx(0.22,0.22,0.53)$ [Fig. 1(a)] \cite{OStockert2004}. Similar dispersive paramagnons up to $E\approx1$~meV were found to stem from $\tau$ in both A-type and S-type CeCu$_2$Si$_2$ \cite{OStockert2011,ZHuesges2018}.
While these dispersive AF fluctuations were discussed in terms of an effective Heisenberg model with short-range magnetic couplings \cite{OStockert2011}, magnetic order in A-type CeCu$_2$Si$_2$ was suggested to result from Fermi surface nesting \cite{OStockert2004}, and AF fluctuations in CeCu$_2$Si$_2$ were found to exhibit critical slowing down consistent with being near a spin-density-wave QCP \cite{JArndt2011}. In the superconducting state of CeCu$_2$Si$_2$, a spin gap forms with spectral weight built up just above it \cite{OStockert2011}, consistent with the formation of a spin resonance mode \ys{with an energy $E_{\rm r}\approx0.2$~meV}, which in the spin-excitonic scenario suggests magnetic pairing \cite{IEremin2008}. The recent discovery of fully-gapped superconductivity in CeCu$_2$Si$_2$ \cite{SKittaka2014,SKittaka2016,TYamashita2017,TTakenaka2017,HIkeda2015,GPang2017,YLi2018,RTazai2018,RTazai2019} challenges the role of magnetic excitations in its superconducting state and calls for a more detailed understanding of its magnetism, including the origin and high-energy properties of its magnetic excitations.  

In this work, by carrying out detailed inelastic neutron scattering measurements over large energy and momentum ranges, we uncover magnetic fluctuations up to $E\gtrsim5$~meV in CeCu$_2$Si$_2$. While magnetic fluctuations below $E\approx1.5$~meV are strongly 3D and dispersive \cite{OStockert2011,JArndt2011}, they become increasingly 2D with increasing energy and form an almost dispersionless column in energy.
By comparing with theoretical calculations, we find magnetic excitations in CeCu$_2$Si$_2$ can be accounted for by intraband scattering of quasiparticles associated with the heavy electron band [Fig. 1(d)],
and therefore allowing us to estimate the band renormalization by matching our calculations with experimental data. We expect this method to be broadly applicable in heavy fermion metals near magnetic criticality. 
The agreement between our experimental and theoretical results suggests \ys{that despite signatures of non-Fermi-liquid behavior \cite{PGegenwart1998,JArndt2011}, the magnetic excitations in the normal state of CeCu$_2$Si$_2$ are reasonably captured by a LDA+$U$ band structure with additional mass renormalization.
}
Our discovery of almost 2D high-energy magnetic excitations in CeCu$_2$Si$_2$ is reminiscent of similar findings in cuprate and iron-based superconductors, 
and favors magnetic pairing with a sign-changing superconducting order parameter.

\section{Results}
\section{Inelastic neutron scattering}
Large single crystals of S-type CeCu$_2$Si$_2$  with $T_{\rm c}\approx0.5$~K were grown using a vertical floating zone method \cite{CCao2011}. 
Multiple crystals with a total mass of $\approx12$~g were co-aligned in the $[H,H,L]$ scattering plane using the E3 four-circle neutron diffractometer at the Chalk River Laboratory \ys{(see Methods section for details)}. Inelastic neutron scattering measurements were carried out using the Multi-Axis Crystal Spectrometer (MACS) \cite{JRodriguez} at the NIST Center for Neutron Research, with fixed outgoing neutron energies $E_{\rm f}=3$ or 5~meV. We reference momentum transfer ${\bf Q}=(H,K,L)$ in reciprocal lattice units (r.l.u.), with $H=a Q_x/(2\pi)$, $K=b Q_y/(2\pi)$, and $L=cQ_z/(2\pi)$ ($a=b\approx4.1$~{\AA} and $c\approx9.9$~{\AA}). Ce$^{3+}$ ions in CeCu$_2$Si$_2$ form a body-centered tetragonal lattice [Fig. 1(b)], and the corresponding Brillouin zone in the $[H,H,L]$ plane is shown in Fig. 1(c). Isotropic background intensities were estimated from regions with $H=0$ and 1, and have been subtracted from our data. 

\begin{figure}[t]
	\includegraphics[scale=.5]{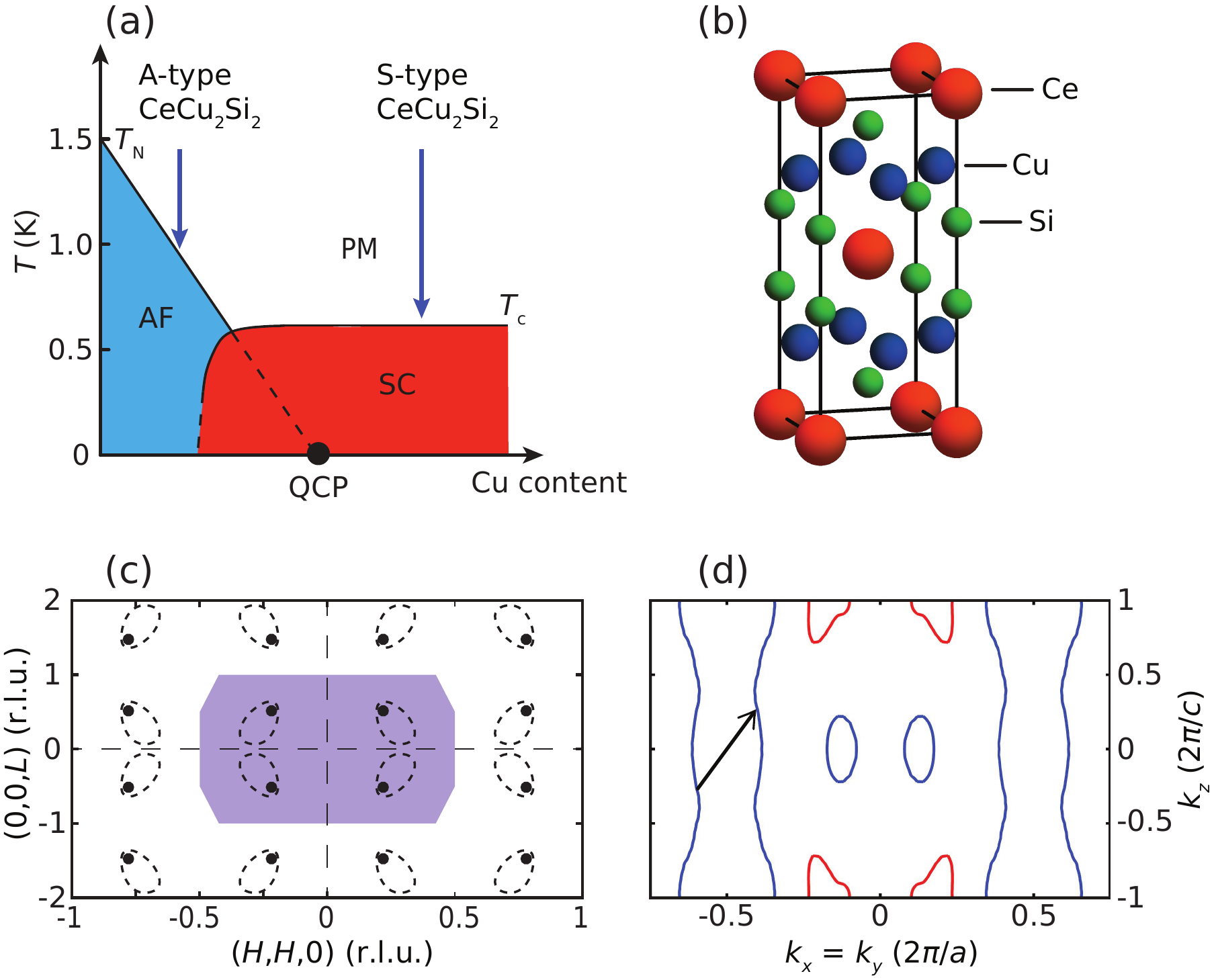}
	\caption{
		(Color online) (a) Schematic phase diagram of CeCu$_2$Si$_2$. (b) Crystal structure of CeCu$_2$Si$_2$, with Ce$^{3+}$ ions forming a body-centered tetragonal lattice. (c) $[H,H,L]$ scattering plane of CeCu$_2$Si$_2$, with the shaded area representing a Brillouin zone for the Ce$^{3+}$ ions. The black dots indicate magnetic Bragg peaks in A-type CeCu$_2$Si$_2$ \cite{OStockert2004}, and the dashed ovals are schematic intensity contours of low-energy AF excitations in our S-type CeCu$_2$Si$_2$. (d) Slice of the CeCu$_2$Si$_2$ Fermi surface with $k_x=k_y$, and the arrow represents the intraband nesting vector. The electronic structure used in this work is identical to that in Ref.~\cite{YLi2018}, \yy{and 3D plots of the Fermi surfaces are shown in Supplementary Fig~7}.
	}
\end{figure}

Maps of the $[H,H,L]$ plane at $T=1.6$~K are compared in the left column of Fig. 2 for different energies, with the corresponding cuts along $(H,H,1.48)$ and $(0.22,0.22,L)$ shown in the middle and right columns. For $E=0.5$~meV, magnetic excitations are relatively sharp, with spectral weight asymmetrically located around $\tau$ [Figs.~2(a)-(c), Fig. 1(c)]. With increasing energy, the magnetic excitations gradually broaden, while maintaining the asymmetric distribution of spectral weight around $\tau$ [Figs.~2(d)-(o)]. Given the same asymmetric distribution is observed for multiple momentum and energy transfers, it is an intrinsic effect rather than a result of instrumental resolution. Two dispersive branches were previously observed at low energies in CeCu$_2$Si$_2$ \cite{OStockert2011}, with the branch closer to the zone boundary being increasingly dominant in intensity with increasing energy \cite{OStockert2011}. In our results a single branch is resolved and can be identified as the dominant branch in previous work, while the weaker branch is unresolved and likely shows up as shoulder in intensity. Clear modulation of magnetic intensity along $(H,H,1.48)$ persists up to the highest measured energy ($E=5.5$~meV), with little or no magnetic intensity at $(0,0,L)$ and $(1,1,L)$ positions; on the other hand, intensity along $(0.22,0.22,L)$ becomes weakly $L$-dependent for $E\geq2$~meV. These observations suggest that although CeCu$_2$Si$_2$ is close to a 3D AF QCP \cite{PGegenwart1998,JArndt2011}, its high-energy magnetic excitations are almost 2D \ys{(see Supplementary Note~1 and Supplementary Fig.~1 for theoretical evidence of quasi-2D magnetic excitations)}, similar to cuprate and iron-based superconductors.

\begin{figure}[t]
	\includegraphics[scale=.21]{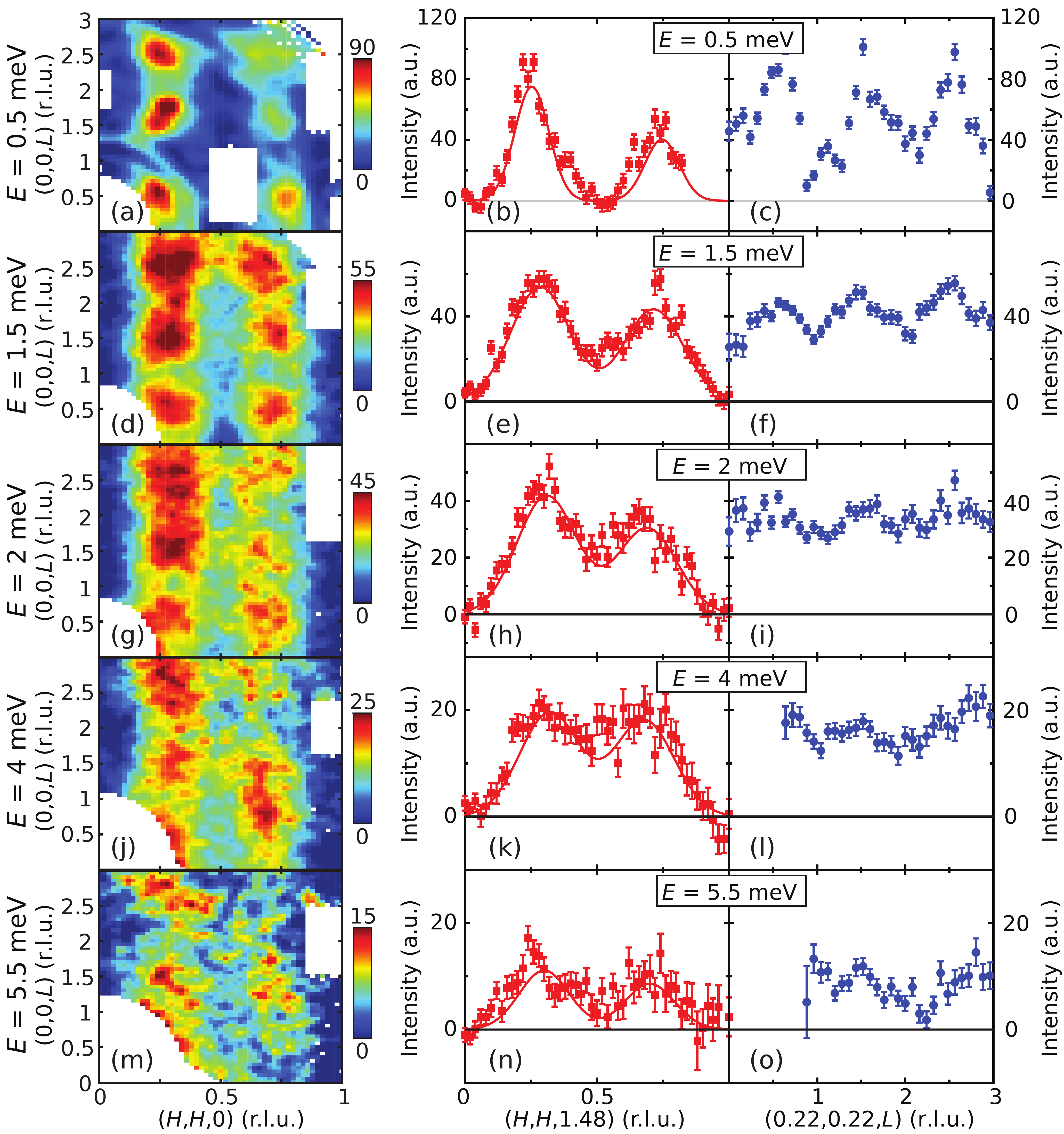}
	\caption{ 
		(Color online) Constant-energy maps of the $[H,H,L]$ plane for different energies are shown in the left panels, corresponding cuts along $(H,H,1.48)$ obtained by binning data with $1.38<L<1.58$ are shown in the middle panels, and corresponding cuts along $(0.22,0.22,L)$ obtained by binning data with $(0.17,0.17)<(H,H)<(0.27,0.27)$ are shown in the right panels. Data in this figure were measured using $E_{\rm f}=5$~meV. All vertical error bars represent statistical errors of 1 s.d.
	}
\end{figure}

Dispersion of the magnetic excitations can be directly visualized in the energy-$(H,H,1.48)$ map in Fig. 3(a).
By fitting scans along $(H,H,1.48)$ using Gaussian peaks symmetrically positioned around (0.5,0.5,1.48) as shown in the middle column of Fig. 2, the magnetic dispersion along $(H,H,1.48)$ can be quantitatively extracted from $E=0.5$~meV to 5.5~meV [Fig. 3(a)] \ys{(see Methods section for details)}. Consistent with previous observations \cite{OStockert2011,JArndt2011,ZHuesges2018}, the magnetic excitations are dispersive for $E\lesssim1.5$~meV, but at higher energies they form a column in energy away from the zone boundary. Such an evolution from dispersive to columnar magnetic excitations is unexpected for a local-moment magnetic system, but have been observed in itinerant magnetic systems, including heavily hole-doped iron pnictides \cite{KHorigane2016}, Fe-doped MnSi$_3$ \cite{STomiyoshi1987} and MnSi \cite{XChen2019}. 

\begin{figure}[t]
	\includegraphics[scale=.33]{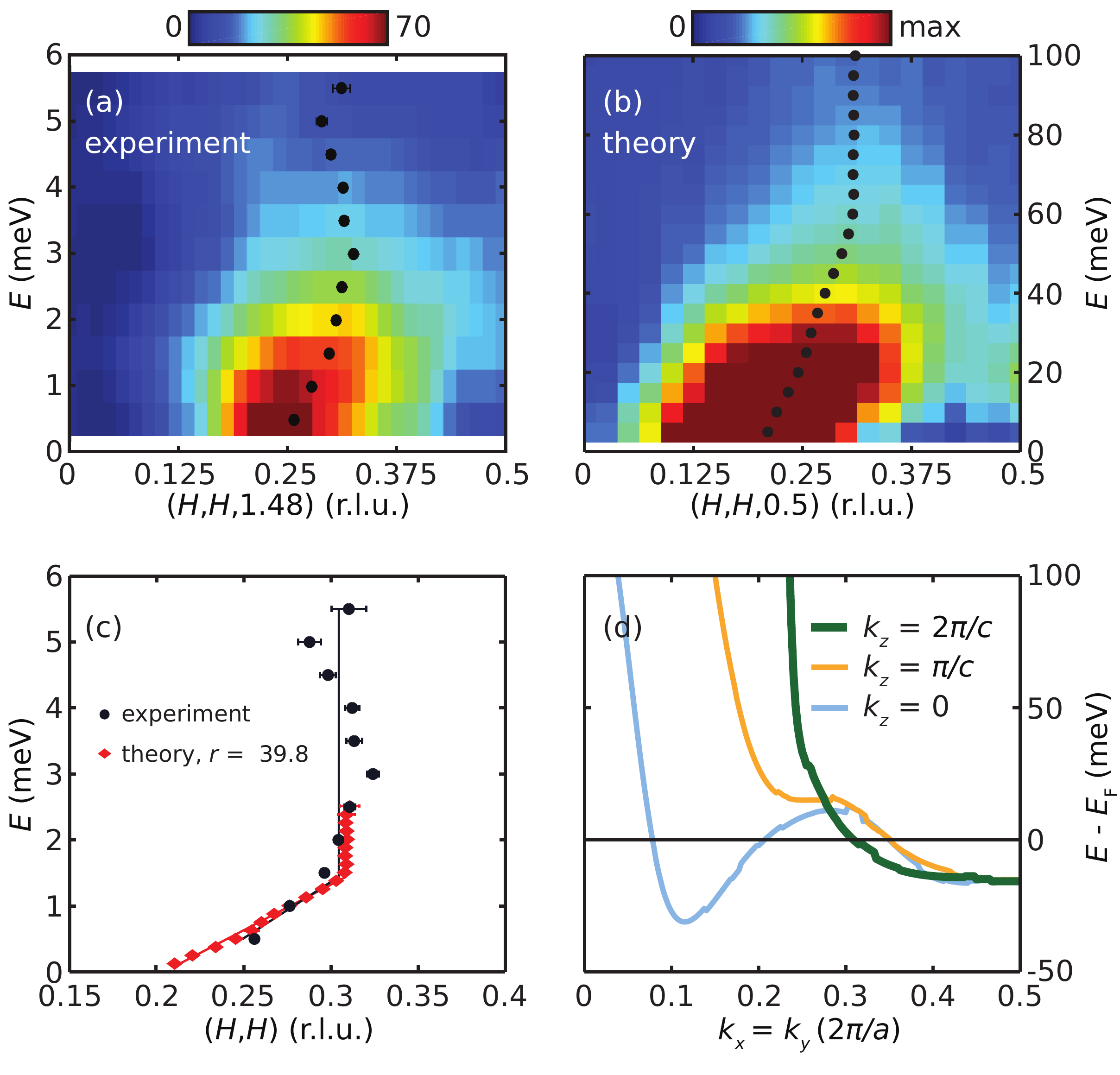}
	\caption{ 
		(Color online) (a) Energy-$(H,H,1.48)$ map at $T=1.6$~K obtained from by binning data with $1.38<L<1.58$. (b) Theoretical energy-$(H,H,0.5)$ maps obtained from RPA calculations using our LDA+$U$ band structure. (c) Comparison of experimental and theoretical dispersions, with the theoretical dispersion scaled down by the band renormalization factor $r$. (d) Dispersion of electron-like bands along $(k_x,k_y=k_x,k_z)$. Data in this figure were measured using $E_{\rm f}=5$~meV. All horizontal error bars are least-square fit errors of 1 s.d.
	}
\end{figure}

\subsection{Theoretical Calculations}
To understand the origin of magnetic excitations in CeCu$_2$Si$_2$ which extend up to at least $E=5.5$~meV, we calculated magnetic excitations within the random-phase approximation (RPA) [Fig.~3(b)] using a LDA+$U$ band structure \yy{(see Methods section for details)}. As can be seen in Fig.~3(b), despite extending over a larger energy scale, the calculated magnetic excitations are in good agreement with our experimental results, with the extracted dispersion consisting of a dispersive part at low energies and a columnar part at high energies \ys{(see Methods section for details)}. By introducing an overall band renormalization factor $r$ that scales our LDA+$U$ band structure, excellent agreement between experimental and theoretical dispersions can be achieved [\ys{Fig.~3(c)}]. We fit the dispersions with a linear-dispersing part that intersects a columnar part at $E_{\rm cross}$, with the experimental dispersion further constrained to stem from $\tau=(0.22,0.22,0.53)$. Scaling the theoretical dispersion so that 
$E_{\rm cross}$ is identical for theoretical and experimental dispersions leads to $r\approx40$. For comparison, the normal state specific heat coefficient from our LDA+$U$ band structure is $C/T\approx50$~mJ/mol$\cdot$K$^2$, which requires $r\approx20$ to match the experimental normal state value of $\approx1.0$~J/mol$\cdot$K$^2$ for $T\rightarrow0$~K \cite{JArndt2011}. 

\begin{figure}[t]
	\includegraphics[scale=.3]{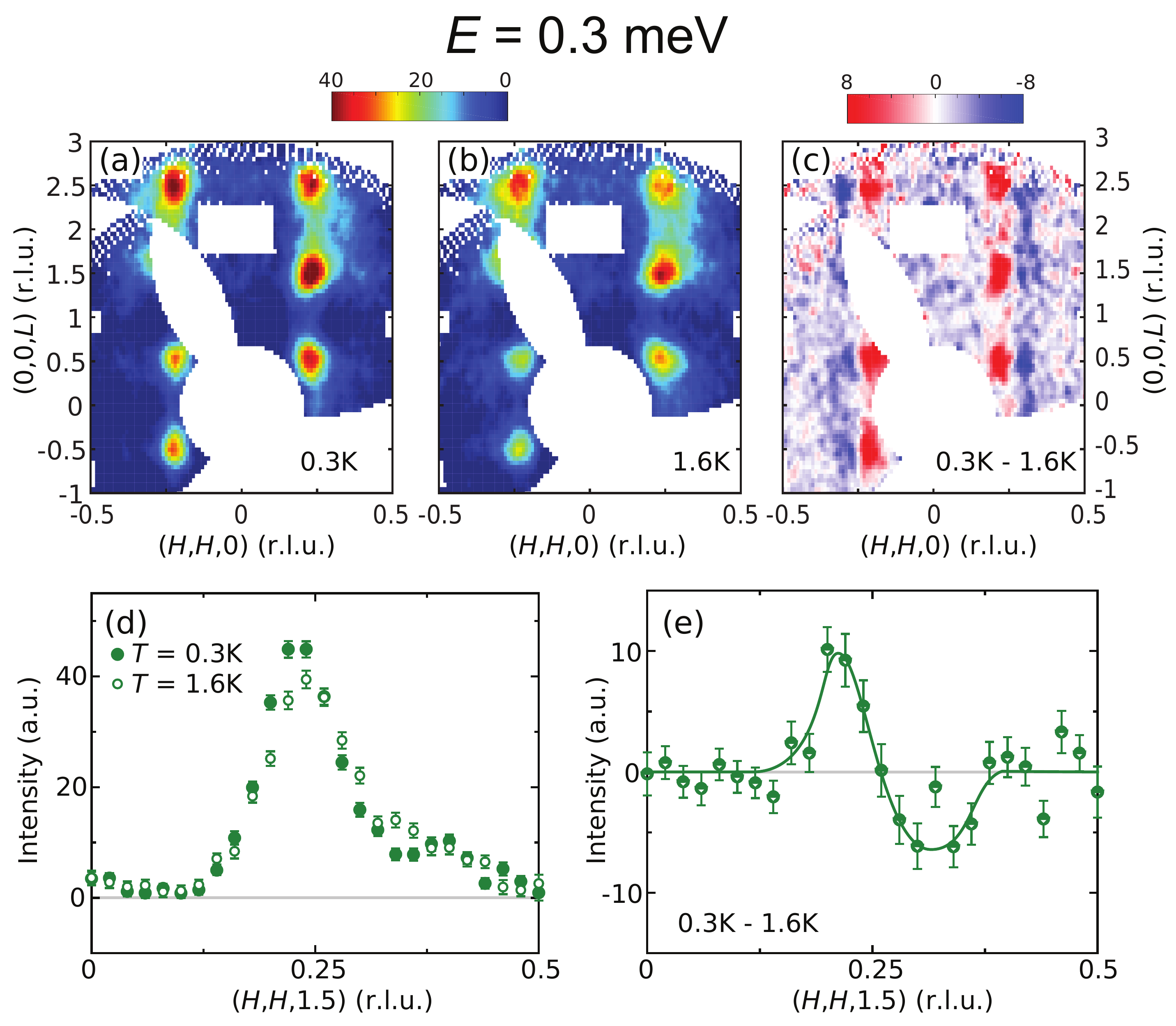}
	\caption{ 
		(Color online) $E=0.3$~meV constant-energy maps of the $[H,H,L]$ plane for (a) $T=0.3$~K, (b) $T=1.6$~K and (c) their difference. (d) Cuts along $(H,H,1.5)$ from maps in (a) and (b), obtained by binning data with $1.4<L<1.6$. (e) The difference of cuts in (d). Data in this figure were measured using $E_{\rm f}=3$~meV. All vertical error bars represent statistical errors of 1 s.d.
	}
\end{figure} 

The Ce $f$ electrons in CeCu$_2$Si$_2$ participate in the formation of an electron and a hole Fermi sheet, both indispensable for its superconducting state \cite{YLi2018}. Compared to specific heat measurements which contain contributions from all Fermi sheets, magnetic excitations measured by neutron scattering are sensitive to bands that exhibit good nesting properties. Therefore, the larger value of $r$ inferred from our neutron scattering results suggests a larger renormalization factor for the well-nested heavy electron band, relative to the hole band. Such band-dependent renormalization effects have been discussed in the context of iron-based superconductors \cite{LdeMedici2014,MYi2017,Zhang14}, with bands exhibiting markedly different renormalization factors for different Fe $3d$ orbitals.
In addition, while specific heat measurements are only sensitive to states within $\sim k_{\rm B}T$ ($\sim0.1$~meV for $T=1.6$~K) of the Fermi level, magnetic fluctuations probed in our experiments involve states on the order several meVs within the Fermi level. Therefore, stronger renormalization effects compared to our LDA+$U$ calculations for states away from the Fermi level (relative to those within $k_{\rm B}T$ of the Fermi level) can also contribute to the larger $r$ values extracted from our inelastic neutron scattering measurements. 
Our findings illustrate the utility of neutron scattering measurements in extracting renormalization factors with band-specificity, complementing specific heat measurements. \ys{It should be noted that such a band-specificity is limited to the well-nested band, which dominates the magnetic excitation spectra, while other bands are effectively not probed.} This method hinges on the fact that magnetic excitations arising from quasiparticles encode information on the band structure \ys{\cite{EAGoremychkin2018,NPButch2015,PYPortnichenko2021}}, and can be especially useful in heavy fermion metals, for which angle-resolved photoemission spectroscopy measurements are challenging due to the small energy scales involved. 

Structure of the heavy electron band along the $(k_x,k_y=k_x,2\pi/c)$ direction [Fig.~3(d)] offers an intuitive understanding of the unusual magnetic dispersion in CeCu$_2$Si$_2$ [Figs.~3(a)-(b)]. The excellent agreement between our experiment and calculations demonstrate that the AF fluctuations in CeCu$_2$Si$_2$ are well-described as particle-hole excitations, in which quasiparticles below the Fermi level are excited to unoccupied states above the Fermi level. In our \ys{LDA+$U$} calculations \ys{(without the renormalization by $r$)}, the band bottom is around $-15$~meV. Therefore, the crossover from dispersive to columnar behavior that occurs $\approx60$~meV is dominated by states above the Fermi level. Comparing states just above the Fermi level ($E\lesssim30$~meV) with those well above the Fermi level ($E>50$~meV) reveals the latter are much lighter, characteristic of conduction bands from none-$f$ orbitals \ys{[Fig.~3(d)]}. Therefore, the crossover from dispersive excitations at low-energies to columnar excitations at high-energies in CeCu$_2$Si$_2$ reflects the electron band's reduction of $f$-orbital content above the Fermi level \yy{(see Supplementary Fig.~2 for the partial density states of CeCu$_2$Si$_2$ from our LDA+$U$ calculations)}. For other values of $k_z$, the band experiences a similar loss of $f$-orbital content for $E>50$~meV, although more complex behaviors are seen closer to the Fermi level. 

\begin{table*}
	\caption{\ys{Published ratios} of the reduction in magnetic exchange energy in the superconducting state $\Delta E_{\rm mag}$ with the superconducting condensation energy $E_{\rm SC}$, for cuprate, iron-based and heavy fermion unconventional superconductors \cite{HWoo2006,CStock2008,OStockert2011,MWang2013,JLeiner2014,SCarr2016}.}
	\begin{ruledtabular}
		\begin{tabular}{ccccccc}
			&YBa$_2$Cu$_3$O$_{6.95}$ &CeCoIn$_5$ &CeCu$_2$Si$_2$ &Ba$_{0.67}$K$_{0.33}$Fe$_2$As$_2$ &Fe$_{1+\delta}$Te$_{1-x}$Se$_x$ ($x\approx0.5$) & NaFe$_{0.9785}$Co$_{0.0215}$As\\
			\hline
			$\frac{\Delta E_{\rm mag}}{\Delta E_{\rm SC}}$&$\approx16$&$\approx35$&$\approx21$&$\approx7$ &$\approx24$&$\approx26$\\
			
		\end{tabular}
	\end{ruledtabular}
	
\end{table*}

\yy{\subsection{Temperature evolution of low-energy magnetic excitations}}

Maps of the $[H,H,L]$ plane for $E=0.3$~meV at $T=0.3$~K ($T<T_{\rm c}$) and 1.6~K ($T>T_{\rm c}$) are compared in Figs.~4(a) and (b), with their difference shown in Fig.~4(c). 
\ys{Since $E = 0.3$~meV is above the energy window of the spin resonance mode in CeCu$_2$Si$_2$ ($E_{\rm r}\approx0.2$~meV) \cite{OStockert2011}, magnetic excitations in the superconducting and normal states are similar. Examining the difference of excitations measured at $T =0.3$~K and 1.6~K nonetheless reveals a subtle shift of magnetic spectral weight towards the Brillouin zone center along $(H,H)$ upon cooling.}
The systematic presence of such a behavior across multiple Brillouin zones [Fig.~4(c)] demonstrates this behavior to be an intrinsic property of CeCu$_2$Si$_2$.
Cuts along the $(H,H,1.5)$ direction are compared in Figs.~4(d) for $T=0.3$ and 1.6~K, and their difference is shown in Fig.~4(e). Such a temperature-dependent shift is similar to the shift of ordering vector \cite{OStockert2004} and magnetic excitations \cite{JArndt2011} in A-type CeCu$_2$Si$_2$, which also move towards the Brillouin zone center upon cooling. 
These observations can be naturally understood now we have shown that magnetic excitations in CeCu$_2$Si$_2$ arise from heavy quasiparticles, and results from a combination of the intrinsic asymmetry of the magnetic dispersion and a depletion of electronic density of states near the Fermi level upon cooling \ys{(see Supplementary Note~2 and Supplementary Fig.~3 for details)}. Such a depletion occurs in A-type CeCu$_2$Si$_2$ due to a spin-density-wave gap, and in S-type CeCu$_2$Si$_2$ due to a superconducting gap. 

\section{Discussion}
Our experimental observation of magnetic excitations extending up to at least $E=5.5$~meV \ys{in CeCu$_2$Si$_2$} demonstrates quasi-2D magnetic fluctuations with an energy scale much larger than the superconducting pairing energy to be a common feature in unconventional superconductors. \ys{As the bandwidth of magnetic excitations is captured by effective magnetic interactions, which in turn determine the saving of magnetic exchange energy in the superconducting state  $\Delta E_{\rm mag}$, the high-energy magnetic excitations observed in our work suggest a $\Delta E_{\rm mag}$ to be at least as large as previous reported \cite{OStockert2011}. The commonality of a much larger $\Delta E_{\rm mag}$ compared to the superconducting condensation energy $\Delta E_{\rm SC}$ [Table I] and the presence of a spin resonance mode \cite{DJScalapino2012,MEschrig2006,GYu2009} across different families of unconventional superconductors, suggest a common pairing mechanism and favors a sign-changing superconducting order parameter such as $d+d$ \cite{GPang2017} or $s^{\pm}$ \cite{HIkeda2015,YLi2018} for CeCu$_2$Si$_2$, rather than $s^{++}$ \cite{RTazai2018}}. To conclusively distinguish between these scenarios, it is important to study dispersion of the spin resonance mode in comparison with theoretical results under different pairing symmetries to test its spin-excitonic nature \cite{YSong2016,YSong2020,WHong2020}. Our work presents a model that captures the normal state magnetic excitations of CeCu$_2$Si$_2$, and is consistent with magnetically driven superconductivity in CeCu$_2$Si$_2$. 

Given that magnetic excitations in  A-type CeCu$_2$Si$_2$ and the superconducting and normal states of S-type CeCu$_2$Si$_2$ are similar for $E\gtrsim0.4$~meV \cite{JArndt2011,ZHuesges2018}, we expect the observed high energy excitations
in our S-type CeCu$_2$Si$_2$ to also be present in A-type CeCu$_2$Si$_2$, as well as compositions in between. In addition, columnar spin excitations near $\tau$ are also seen in CeNi$_2$Ge$_2$, although compared to CeCu$_2$Si$_2$ no low-energy dispersive features were reported \cite{BFak2000}. Since CeNi$_2$Ge$_2$ is paramagnetic and relatively far away from an AF QCP, this suggests that compared to the low-energy dispersive excitations, the columnar excitations at high energies are more robust upon tuning away from the QCP.
In both cuprate and iron-based superconductors, the quasi-2D high energy magnetic excitations remain robust when tuning towards the superconducting state, while the low-energy excitations may change dramatically. Therefore, the prevalence of robust high-energy magnetic excitations suggest short-range 2D magnetic correlations provide a backdrop from which unconventional superconductivity emerges, while low-energy AF fluctuations and electronic structure can range from strongly 2D to having significant 3D features and are more tunable. This is analogous to conventional superconductors, in which a large Debye cutoff energy provides a backdrop for potential high-temperature superconductivity.  

\section{Methods}
\subsection{Sample preparation \yy{and inelastic neutron scattering measurements}}
Several rod-shaped S-type CeCu$_2$Si$_2$ single crystal samples were grown using a vertical optical heating floating zone method [Supplementary Fig.~4(a)] \cite{CCao2011}. To avoid Si excess which results in A-type CeCu$_2$Si$_2$, we used a high-pressure Ar atmosphere and a relatively small overheating of the floating zone beyond the melting temperature, which effectively reduces Cu evaporation. Using inductively coupled plasma atomic-emission spectroscopy, we determined the atomic percentage of our samples to be Ce:20.1(1)\%, Cu:40.3(1)\% and Si:39.6(1)\%. This stoichiometry is found to be consistent across several pieces of our samples, indicating they are dominantly S-type \cite{FSteglich1996} in agreement with previous transport measurements \cite{CCao2011}. \ys{Specific heat measurements were carried out for several pieces of our CeCu$_2$Si$_2$ samples [Supplementary Fig.~4(c)], all exhibiting a specific heat jump below $T_{\rm c}\approx0.5$~K, different from A-type and A/S-type CeCu$_2$Si$_2$ samples \cite{OStockert2004,ELengyel2011}. Magnitude of the specific heat jump exhibits some sample-dependence, possibly due to parts of the samples being nonsuperconducting. The specific heat measurements evidence our CeCu$_2$Si$_2$ samples are dominantly S-type, without prominent signatures of antiferromagnetism.} While the presence of a minority phase of A-type or A/S-type CeCu$_2$Si$_2$ \ys{is difficult to rule out}, the high-energy magnetic excitations uncovered in our work should also be present in A-type and A/S-type CeCu$_2$Si$_2$, \ys{as discussed above}. Therefore, the possible presence of such minority phases will not affect the conclusions of our work. 

We cut the rod-like samples into segments of a few centimeters, and used the E3 four-circle neutron diffractometer to identify single-grain pieces by mapping the $\phi$ and $\chi$ rotation angles, with the scattering angle $2\theta$ adjusted to the scattering angle of an intense structural Bragg peak. We then co-aligned four such segments in the $[H,H,L]$ plane, as shown in Supplementary Fig.~4(b).

Inelastic neutron scattering measurements were carried out using the Multi-Axis Crystal Spectrometer (MACS) \cite{JRodriguez} at the NIST center for neutron research (NCNR) in Gaithersburg, MD. Our measurements were carried out using fixed $E_{\rm f}=3$ or 5~meV, with Be filters placed after the sample for both $E_{\rm f}$ and before the sample for $E_{\rm f}=3$~meV. MACS consists of 20 spectroscopic detectors, and by rotating the sample and shifting the detectors, a map of the scattering plane at a fixed energy transfer can be efficiently constructed. The double-bounce analyzers are vertically focused, while the monochromator is doubly focused. Instrumental energy resolutions at the elastic line are $\Delta E\approx0.14$~meV for $E_{\rm f}=3$~meV, and $\Delta E\approx0.35$~meV for $E_{\rm f}=5$~meV. Sample alignment is confirmed on MACS for the (110) and (002) structural Bragg peaks. As shown in Supplementary Fig.~5, our samples are reasonably well-aligned for sample arrays used in inelastic neutron scattering measurements. 

\subsection{Extraction of experimental and calculated dispersions}
To extract the experimental dispersion of magnetic excitations in CeCu$_2$Si$_2$, cuts along $(H,H,1.48)$ were obtained by binning data with $1.38<L<1.58$ and fit using two Gaussian peaks

\begin{align}	
	I(x)=a_1\exp(-\frac{(x-\delta)^2}{2c^2})+a_2\exp(-\frac{(x-1+\delta)^2}{2c^2}).	
\end{align}

The same expression with an additional constant term is used to extract the calculated dispersion. In these fittings, $x=0$ corresponds to $(0,0)$, $x=1$ corresponds to $(1,1)$ and $\delta$ is the fit peak center position. Representative fits to our experimental data are shown in \ys{the middle column of Fig.~3}, and representative fits to our theoretical results are shown in Supplementary Fig.~6.



\yy{
\subsection{LDA+$U$ band structure}
The LDA+$U$ band structure calculations were performed using the full-potential augmented plane-wave plus local orbital method as implemented in WIEN2k \cite{WIEN2k}. The Perdew-Burke-Ernzerhof exchange-correlation energy \cite{PBE} was used with spin-orbit coupling and an effective on-site Coulomb interaction $U = 5$~eV \cite{VIAnisimov}. The orbital characters were obtained using WANNIER90 code \cite{AAMostofi} via WIEN2WANNIER interface \cite{JKunes}. Our LDA+$U$ band structure was used previously to study the pairing symmetry of CeCu$_2$Si$_2$ \cite{YLi2018}, and is similar to band structures in previous LDA+$U$ calculations \cite{SKittaka2014,HIkeda2015} and from the renormalized band approach \cite{GZwicknagl1993}. 
As the $f$-electrons are itinerant in our LDA+$U$ calculations, the obtained Fermi surfaces are ``large".
}

\yy{
We note that there is a subtle difference in the band structures of Refs.~\cite{HIkeda2015} and \cite{YLi2018}, with the latter used in the calculations of this work. Comparing the heavy electron Fermi surfaces in these two works, there is an extra ring-like Fermi surface in Refs.~\cite{YLi2018} around the $\Gamma$ point (Supplementary Fig.~7). This difference results from details in implementing the calculations, and affects neither key features of the band structure nor the expected physics, which is dominated by the cylindrical heavy electron Fermi surface common to Refs.~\cite{HIkeda2015} and \cite{YLi2018}.
}

\yy{
The partial density of states of CeCu$_2$Si$_2$ from our LDA+$U$ calculations is shown in Supplementary Fig.~2. As can be seen, the Ce-$f_{5/2}$ density of states is mainly located just above the Fermi level, and decreases rapidly above 50~meV, becoming increasingly small around 100~meV.  Such an evolution of the partial density of states is consistent with the notion that a change of character of the band states causes the crossover from dispersive to columnar magnetic excitations. However, as the partial density of states contains contributions from all the electronic states, and not just the well-nested regions that give rise to the magnetic excitations, the signatures for such a change is not as clear in the partial density of states compared to Fig.~3(d). 
}

\subsection{Calculation of magnetic excitations in CeCu$_2$Si$_2$}
The bare magnetic susceptibility with four indices is:
\begin{align}
	[\chi_{0}]^{\mu\mu'}_{\nu\nu'}(\textbf{q},i\omega_{n}) = & -  \sum_{\textbf{k},n_{1},n_{2}} a^{\nu}_{n_{1}}(\textbf{k})a^{\nu'*}_{n_{1}}(\textbf{k}) a^{\mu'}_{n_{2}}(\textbf{k}+\textbf{q})a^{\mu*}_{n_{2}}(\textbf{k}+\textbf{q}) \\
	& \times { 1 \over \beta} \sum_{i\omega_{m}} {1 \over i\omega_{m}-\epsilon_{n_{1}}(\textbf{k})} {1 \over i\omega_{m}-i\omega_{n}-\epsilon_{n_{2}}(\textbf{k}+\textbf{q})}
\end{align}

where $a^{\mu}_{n}(\textbf{k})$, $\epsilon_{n}(\textbf{k})$ are the unitary matrices diagonalizing $H_{0}$ and the energy dispersion,  respectively. The sum over $n$ is taken over the entire band index.
Using the Matsubara frequency sum rule and the Fermi-Dirac function $n_{F}(\epsilon)={1 \over e^{\beta \epsilon}+ 1}$,  we get:
\begin{align}
	[\chi_{0}]^{\mu\mu'}_{\nu\nu'}(\textbf{q},i\omega_{n}) 
	= -  \sum_{\textbf{k},n_{1},n_{2}} a^{\nu}_{n_{1}}(\textbf{k})a^{\nu'*}_{n_{1}}(\textbf{k}) a^{\mu'}_{n_{2}}(\textbf{k}+\textbf{q})a^{\mu*}_{n_{2}}(\textbf{k}+\textbf{q}) 
	\times {n_{F}\left[\epsilon_{n_{2}}(\textbf{k}+\textbf{q})\right] - n_{F}\left[\epsilon_{n_{1}}(\textbf{k})\right] \over i\omega_{n} + \epsilon_{n_{2}}(\textbf{k}+\textbf{q})-\epsilon_{n_{1}}(\textbf{k})}
\end{align}
using $i\omega \rightarrow \omega+i\eta$, we have
\begin{align}
	[\chi_{0}]^{\mu\mu'}_{\nu\nu'}(\textbf{q},\omega) = & -  \sum_{\textbf{k},n_{1},n_{2}} a^{\nu}_{n_{1}}(\textbf{k})a^{\nu'*}_{n_{1}}(\textbf{k}) a^{\mu'}_{n_{2}}(\textbf{k}+\textbf{q})a^{\mu*}_{n_{2}}(\textbf{k}+\textbf{q}) 
	\times {n_{F}\left[\epsilon_{n_{2}}(\textbf{k}+\textbf{q})\right] - n_{F}\left[\epsilon_{n_{1}}(\textbf{k})\right]\over \omega + \epsilon_{n_{2}}(\textbf{k}+\textbf{q})-\epsilon_{n_{1}}(\textbf{k}) + i\eta}
\end{align}

The transverse RPA susceptibility for a multiband system is: 
\begin{equation}
	\hat{\chi}_{\rm RPA} = \frac{\hat{\chi}_0}{1 - \hat U \hat{\chi}_0 }
\end{equation}
This is in fact a Bethe-Salpeter equation where $\hat{U}$ is a $n^2\times n^2$ matrix with orbital number $n$.
\begin{equation}
	\hat U^{rr}_{rr} = U ,\ 
	\hat U^{rr}_{ss} = U' ,\ 
	\hat U^{rs}_{rs} = J ,\ 
	\hat U^{rs}_{sr} = J' \ (r\not=s).
\end{equation}


On-site interactions $U=U'$=0.25~eV\yy{, $J=J'=0$~eV and a $20\times20\times20$ $k$-mesh} were used in our RPA calculations, similar to previous work \cite{HIkeda2015}.  The $U$ values used in our RPA calculations are smaller than those in our LDA+$U$ calculations to avoid the divergence of RPA magnetic susceptibility. The key features of the RPA susceptibility are mostly determined by the bare susceptibility $\chi_0$ \yy{(Supplementary Fig.~8), which already contains the essential features of the magnetic susceptibility,} and are not strongly affected by the interaction term $U$.

{\bf Disclaimer:} The identification of any commercial product or trade name does not imply endorsement or recommendation by the National Institute of Standards and Technology.

\section{Data Availability}
The data that support the findings of this study are available from the corresponding authors upon reasonable request.

\section{Competing Interests}
The authors declare no competing interests.

\section{Contributions}
P. D., Y. Song, and C. C. conceived the project. C. C. and W. L. prepared the samples. Z. Y. and Y. Song co-aligned the samples. W. W., Y. Song, C. C. and Y. Q. carried out the experiments. W. W. and Y. Song analyzed the data. Y. X., Y. Sheng and Y. Y. carried out the theoretical analyses. Y. Song, P. D. and Y. Y. wrote the paper with input from all authors.    

\section{Acknowledgements}
\ys{We would like to thank Binod K. Rai and Emilia Morosan for help with specific heat measurements.}
Neutron scattering work at Rice is supported by the U.S. Department of Energy, BES under Grant No. DE-SC0012311. The single-crystal characterization work at Rice is supported by Robert A. Welch Foundation Grant No. C-1839. The work at NWPU is supported by National Key Research and Development Program of China (2016YFB1100101), National Natural Science Foundation of China ( 51971180,51971179), Key Research and Development Program of Shaanxi (2021KWZ-13) and Guangdong Science and Technology program (2019B090905009). The work at Lawrence Berkeley National Laboratory and the University of California, Berkeley is supported by the Office of Science, Office of Basic Energy Sciences (BES), Materials Sciences and Engineering Division of the U.S. Department of Energy (DOE) under Contract No. DE-AC02-05-CH11231 within the Quantum Materials Program (KC2202) and BES, U.S. DOE Grant No. DE-AC03-76SF008. The work at IOP, Beijing is supported by the National Key R\&D Program of China (2017YFA0303103) and the National Natural Science Foundation of China (11974397, 11774401). Access to MACS was provided by the Center for High Resolution Neutron Scattering, a partnership between the National Institute of Standards and Technology and the National Science Foundation under Agreement No. DMR-1508249.


\begin{thebibliography}{}
\bibitem{FSteglich} F. Steglich, J. Aarts, C. D. Bredl, W. Lieke, D. Meschede, W. Franz, and H. Sch\"{a}fer, Phys. Rev. Lett. {\bf 43}, 1892 (1979).

\bibitem{DJScalapino2012} D. J. Scalapino, Rev. Mod. Phys. {\bf 84}, 1383 (2012).

\bibitem{GStewart2017} G. R. Stewart, Adv. Phys. {\bf 66}, 75-196 (2017). 	

\bibitem{PALee2006} P. A. Lee, N. Nagaosa, and X. G. Wen, Rev. Mod. Phys. {\bf 78}, 17 (2006).

\bibitem{BKeimer2015} B. Keimer, S. A. Kivelson, M. R. Norman, S. Uchida, and J. Zaanen, Nature {\bf 518}, 179-186 (2015).

\bibitem{Sidis} Y. Sidis, S. Pailh\`{e}s, V. Hinkov, B. Fauqu\'{e}, C. Ulrich,
L. Capogna, A. Ivanov, L. P. Regnault, B. Keimer, P. Bourges, C. R. Physique {\bf 8}, 745 (2007).

\bibitem{GStewart2011} G. R. Stewart, Rev. Mod. Phys. {\bf 83}, 1589 (2011). 	

\bibitem{PDai2015} P. Dai, Rev. Mod. Phys. {\bf 87}, 855 (2015).

\bibitem{RMFernandes2014} R. M. Fernandes, A. V. Chubukov, and J. Schmalian, Nat. Phys. {\bf 10}, 97-104 (2014).

\bibitem{QMSi2016} Q. Si, R. Yu, and E. Abrahams, Nat. Rev. Mater. {\bf 1} 16017 (2016).

\bibitem{CPfleiderer2009} C. Pfleiderer, Rev. Mod. Phys. {\bf 81}, 1551 (2009).

\bibitem{OStockert2012} O. Stockert, S. Kirchner, F. Steglich, and Q. Si, J. Phys. Soc. Jpn. {\bf 81}, 011001 (2012).

\bibitem{BDWhite2015} B. D. White, J. D. Thompson, and M. B. Maple, Physica C {\bf 514}, 246-278 (2015).

\bibitem{TMoriya2003} T. Moriya and K. Ueda, Rep. Prog. Phys. {\bf 66}, 1299-1341 (2003).

\bibitem{EDemler1998} E. Demler and S. C. Zhang, Nature {\bf 396}, 733-735 (1998).

\bibitem{HWoo2006} H. Woo, P. Dai, S. M. Hayden, H. A. Mook, T. Dahm, D. J. Scalapino, T. G. Perring, and F. Do\u{g}an, Nat. Phys. {\bf 2} 600-604 (2006). 

\bibitem{CStock2008} C. Stock, C. Broholm, J. Hudis, H. J. Kang, and C. Petrovic, Phys. Rev. Lett. {\bf 100}, 087001 (2008).

\bibitem{OStockert2011} O. Stockert, J. Arndt, E. Faulhaber, C. Geibel, H. S. Jeevan, S. Kirchner, M. Loewenhaupt, K. Schmalzl, W. Schmidt, Q. Si, and F. Steglich, Nat. Phys. {\bf 7}, 119-124 (2011).

\bibitem{MWang2013} M. Wang, C. Zhang, X. Lu, G. Tan, H. Luo, Y. Song, M. Wang, X. Zhang, E.A. Goremychkin, T.G. Perring, T.A. Maier, Z. Yin, K. Haule, G. Kotliar, and P. Dai, Nat. Commun. {\bf 4}, 2874 (2013). 

\bibitem{JLeiner2014} J. Leiner, V. Thampy, A. D. Christianson, D. L. Abernathy, M. B. Stone, M. D. Lumsden, A. S. Sefat, B. C. Sales, Jin Hu, Zhiqiang Mao, Wei Bao, and C. Broholm, Phys. Rev. B {\bf 90}, 100501(R) (2014).

\bibitem{SCarr2016} S. V. Carr, C. Zhang, Y. Song, G. Tan, Yu Li, D. L. Abernathy, M. B. Stone, G. E. Granroth, T. G. Perring, and P. Dai, Phys. Rev. B {\bf 93}, 214506 (2016).

\bibitem{TDahm2009} T. Dahm, V. Hinkov, S. V. Borisenko, A. A. Kordyuk, V. B. Zabolotnyy, J. Fink, B. B\"{u}chner,
D. J. Scalapino, W. Hank, and B. Keimer, Nat. Phys. {\bf 5}, 217-221 (2009).

\bibitem{MEschrig2006} M. Eschrig, Adv. Phys. {\bf 55}, 47-183 (2006).

\bibitem{GYu2009} G. Yu, Y. Li, E. M. Motoyama, and M. Greven, Nat. Phys. {\bf 5}, 873 (2009).

\bibitem{weber2019} F. Wa$\rm \beta$er, J. T. Park, S. Aswartham, S. Wurmehl, Y. Sidis, P. Steffens, K. Schmalzl, B. B$\rm \ddot{u}$chner, and Markus Braden, 
npj Quantum Materials {\bf 4}, 59 (2019). 

\bibitem{RJBirgeneau2006} R. J. Birgeneau, C. Stock, J. M. Tranquada, and K. Yamada, J. Phys. Soc. Jpn. {\bf 75}, 111003 (2006).

\bibitem{OJLipscombe2007} O. J. Lipscombe, S. M. Hayden, B. Vignolle, D. F. McMorrow, and T. G. Perring, Phys. Rev. Lett. {\bf 99}, 067002 (2007).

\bibitem{MFujita2012} M. Fujita, H. Hiraka, M. Matsuda, M. Matsuura, J. M. Tranquada, S. Wakimoto, G. Xu, and K. Yamada, J. Phys. Soc. Jpn. {\bf 81}, 011007 (2012).

\bibitem{MLeTacon2011} M. Le Tacon, G. Ghiringhelli, J. Chaloupka, M. Moretti Sala, V. Hinkov, M. W. Haverkort, M. Minola, M. Bakr, K. J. Zhou, S. Blanco-Canosa, C. Monney, Y. T. Song, G. L. Sun, C. T. Lin, G. M. De Luca, M. Salluzzo, G. Khaliullin, T. Schmitt, L. Braicovich, and B. Keimer, Nat. Phys. {\bf 7}, 725-730 (2011).

\bibitem{MLeTacon2013} M. Le Tacon, M. Minola, D. C. Peets, M. Moretti Sala, S. Blanco-Canosa, V. Hinkov, R. Liang, D. A. Bonn, W. N. Hardy, C. T. Lin, T. Schmitt, L. Braicovich, G. Ghiringhelli, and B. Keimer, Phys. Rev. B {\bf 88}, 020501(R) (2013).

\bibitem{MPMDean2013} M. P. M. Dean, G. Dellea, R. S. Springell, F. Yakhou-Harris, K. Kummer, N. B. Brookes, X. Liu, Y-J. Sun, J. Strle, T. Schmitt, L. Braicovich, G. Ghiringhelli, I. Bo\v{z}ovi\'{c}, and J. P. Hill, Nat. Mater. {\bf 12}, 1019-1023 (2013).

\bibitem{MLiu2012} M. Liu, L. W. Harriger, H. Luo, M. Wang, R. A. Ewings, T. Guidi, H. Park, K. Haule, G. Kotliar, S. M. Hayden, and P. Dai, Nat. Phys. {\bf 8}, 376-381 (2012).

\bibitem{KJZhou2013} K. J. Zhou, Y. B. Huang, C. Monney, X. Dai, V. N. Strocov, N. L. Wang, Z. G. Chen, C. Zhang, P. Dai, L. Patthey, J. van den Brink, H. Ding, T. Schmitt, Nat. Commun. {\bf 4}, 1470 (2013).







\bibitem{PGegenwart1998} P. Gegenwart, C. Langhammer, C. Geibel, R. Helfrich, M. Lang, G. Sparn, F. Steglich, R. Horn, L. Donnevert, A. Link, and W. Assmus, Phys. Rev. Lett. {\bf 81}, 1501 (1998).	

\bibitem{JArndt2011} J. Arndt, O. Stockert, K. Schmalzl, E. Faulhaber, H. S. Jeevan, C. Geibel, W. Schmidt, M. Loewenhaupt, and F. Steglich, Phys. Rev. Lett. {\bf 106}, 246401 (2011).


\bibitem{OStockert2004} O. Stockert, E. Faulhaber, G. Zwicknagl, N. St\"{u}{\ss}er,4 H. S. Jeevan, M. Deppe, R. Borth, R. K\"{u}chler, M. Loewenhaupt, C. Geibel, and F. Steglich, Phys. Rev. Lett. {\bf 92}, 136401 (2004).

\bibitem{ZHuesges2018} Z. Huesges, K. Schmalzl, C. Geibel, M. Brando, F. Steglich, and O. Stockert, Phys. Rev. B {\bf 98}, 134425 (2018).




\bibitem{IEremin2008} I. Eremin, G. Zwicknagl, P. Thalmeier, and P. Fulde, Phys. Rev. Lett. {\bf 101}, 187001 (2008).

\bibitem{SKittaka2014} S. Kittaka, Y. Aoki, Y. Shimura, T. Sakakibara, S. Seiro, C. Geibel, F. Steglich, H. Ikeda, and K. Machida, Phys. Rev. Lett. {\bf 112}, 067002 (2014).

\bibitem{TYamashita2017} T. Yamashita, T. Takenaka, Y. Tokiwa, J. A. Wilcox, Y. Mizukami, D. Terazawa, Y. Kasahara, S. Kittaka, T. Sakakibara, M. Konczykowski, S. Seiro, H. S. Jeevan, C. Geibel, C. Putzke, T. Onishi, H. Ikeda, A. Carrington, T. Shibauchi, and Y. Matsuda, Sci. Adv. {\bf 3}, e1601667 (2017).

\bibitem{TTakenaka2017} T. Takenaka, Y. Mizukami, J. A. Wilcox, M. Konczykowski, S. Seiro, C. Geibel, Y. Tokiwa, Y. Kasahara, C. Putzke, Y. Matsuda, A. Carrington, and T. Shibauchi, Phys. Rev. Lett. {\bf 119}, 077001 (2017).

\bibitem{SKittaka2016} S. Kittaka, Y. Aoki, Y. Shimura, T. Sakakibara, S. Seiro, C. Geibel, F. Steglich, Y. Tsutsumi, H. Ikeda, and K. Machida, Phys. Rev. B {\bf 94}, 054514 (2016).

\bibitem{GPang2017} G. Pang, M. Smidman, J. Zhang, L. Jiao, Z. Weng, E. M. Nica, Y. Chen, W. Jiang, Y. Zhang, W. Xie, H. S. Jeevan, H. Lee, P. Gegenwart, F. Steglich, Q. Si, and H. Yuan, Proc. Natl. Acad. Sci. U.S.A. {\bf 115} 5343-5347 (2017).

\bibitem{HIkeda2015} H. Ikeda, M. T. Suzuki, and R. Arita, Phys. Rev. Lett. {\bf 114}, 147003 (2015). 

\bibitem{YLi2018} Y. Li, M. Liu, Z. Fu, X. Chen, F. Yang, and Y. Yang, Phys. Rev. Lett. {\bf 120}, 217001 (2018).

\bibitem{RTazai2018} R. Tazai and H. Kontani, Phys. Rev. B {\bf 98}, 205107 (2018).

\bibitem{RTazai2019} R. Tazai and H. Kontani, J. Phys. Soc. Jpn. {\bf 88}, 063701 (2019).

\bibitem{CCao2011} C. D. Cao, M. Deppe, G. Behr, W. L\"{o}ser, N. Wizent, O. Kataeva, and B. B\"{u}chner, Cryst. Growth Des. {\bf 11}, 431-435 (2011).

\bibitem{JRodriguez} J. A. Rodriguez, D. M. Adler, P. C. Brand, C. Broholm, J. C. Cook, C. Brocker, R. Hammond, Z. Huang, P. Hundertmark, J. W. Lynn, N. C. Maliszewskyj, J. Moyer, J. Orndorff, D. Pierce, T. D. Pike, G. Scharfstein, S. A. Smee and R. Vilaseca, Meas. Sci. Technol. {\bf 19}, 034023 (2008).

\bibitem{KHorigane2016} K. Horigane, K. Kihou, K. Fujita, R. Kajimoto, K. Ikeuchi, S. Ji, J. Akimitsu, and C. H. Lee, Sci. Rep. {\bf 6}, 33303 (2016).

\bibitem{STomiyoshi1987} S. Tomiyoshi, Y. Yamaguchi, M. Ohashi, E. R. Cowley, and G. Shirane, Phys. Rev. B {\bf 36}, 2181 (1987).

\bibitem{XChen2019} X. Chen, I. Krivenko, M. B. Stone, A. I. Kolesnikov, T. Wolf, D. Reznik, K. S. Bedell, F. Lechermann, and S. D. Wilson, Nat. Commun. {\bf 11}, 3076 (2020).



\bibitem{LdeMedici2014} L. de' Medici, G. Giovannetti, and M. Capone, Phys. Rev. Lett. {\bf 112}, 177001 (2014).

\bibitem{MYi2017} M. Yi, Y. Zhang, Z. X. Shen, and D. Lu, npj Quant. Mater. {\bf 2}, 57 (2017).

\bibitem{Zhang14} C. Zhang, L. W. Harriger, Z. Yin, W. Lv, M. Wang, G. Tan, Y. Song, D. L. Abernathy, W. Tian, T. Egami, K. Haule, G. Kotliar, and P. Dai, Phys. Rev. Lett. {\bf 112}, 217202 (2014).

\bibitem{EAGoremychkin2018} E. A. Goremychkin, H. Park, R. Osborn, S. Rosenkranz, J. P. Castellan, V. R. Fanelli, A. D. Christianson, M. B. Stone, E. D. Bauer, K. J. McClellan, D. D. Byler, and J. M. Lawrence, Science {\bf 359}, 186-191 (2018).

\bibitem{NPButch2015}
Nicholas P. Butch, Michael E. Manley, Jason R. Jeffries, Marc Janoschek, Kevin Huang, M. Brian Maple, Ayman H. Said, Bogdan M. Leu, and Jeffrey W. Lynn, Phys. Rev. B {\bf 91}, 035128 (2015).

\bibitem{PYPortnichenko2021} P. Y. Portnichenko, A. S. Cameron, and D. S. Inosov, "Neutron-Scattering Studies of Spin Dynamics in Pure and Doped CeB$_6$", Rare Earth Borides, edited by D. S. Inosov, Jenny Stanford Publishing (2021). 

\bibitem{YSong2016} Y. Song, J. Van Dyke, I.K. Lum, B.D. White, S. Jang, D. Yazici, L. Shu, A. Schneidewind, P. \v{C}erm\'{a}k, Y. Qiu, M.B. Maple, D. K. Morr, P. Dai, Nat. Commun. {\bf 7}, 12774 (2016).	

\bibitem{YSong2020} Y. Song, W. Wang, J. S. V. Dyke, N. Pouse, S. Ran, D. Yazici, A. Schneidewind, P. \v{C}erm\'{a}k, Y. Qiu, M. B. Maple, D. K. Morr, P. Dai, Commun. Phys. {\bf 3}, 98 (2020). https://doi.org/10.1038/s42005-020-0365-8


\bibitem{WHong2020} W. Hong, L. Song, B. Liu, Z. Li, Z. Zeng, Y. Li, D. Wu, Q. Sui, T. Xie, S. Danilkin, H. Ghosh, A. Ghosh, J. Hu, L. Zhao, X. Zhou, X. Qiu, S. Li, and H. Luo, Phys. Rev. Lett. {\bf 125}, 117002 (2020).

\bibitem{BFak2000} B. F\r{a}k, J. Flouquet, G. Lapertot, T. Fukuhara, and H. Kadowaki, J. Phys.: Condens. Matter {\bf 12}, 5423-5435 (2000).


\bibitem{FSteglich1996} F. Steglich, P. Gegenwart, C. Geibel, R. Helfrich, P. Hellmann, M. Lang, A. Link, R. Modler, G. Sparn, N. B\"{u}ttgen, A. Loidl, Physica B {\bf 223-224}, 1-8 (1996).

\bibitem{ELengyel2011} E. Lengyel, M. Nicklas, H. S. Jeevan, C. Geibel, and F. Steglich, Phys. Rev. Lett. {\bf 107}, 057001 (2011)

\bibitem{WIEN2k} P. Blaha, K. Schwarz, G. K. H. Madsen, D. Kvasnicka, and J. Luitz, {\it WIEN2k: An Augmented Plane Wave + Local Orbitals Program for Calculating Crystal Properties} (Karlheinz Schwarz, Technische Universit\"{a}t Wien, 2014).

\bibitem{PBE} J. P. Perdew, K. Burke, and M. Ernzerhof, Phys. Rev. Lett. {\bf 77}, 3865 (1996).

\bibitem{VIAnisimov} V. I. Anisimov, F. Aryasetiawa, and A. I. Lichtenstein, J. Phys.: Condens. Matter {\bf 9}, 767 (1997).

\bibitem{AAMostofi} A. A. Mostofi, J. R. Yates, Y.-S. Lee, I. Souza, D. Vanderbilt, and N. Marzari, Comput. Phys. Commun. {\bf 178}, 685 (2008).

\bibitem{JKunes} J. Kune\v{s}, R. Arita, P. Wissgott, A. Toschi, H. Ikeda, and K. Held, Comput. Phys. Commun. {\bf 181}, 1888 (2010).

\bibitem{GZwicknagl1993} Gertrud Zwicknagl and Uwe Pulst, Physica B {\bf 186-188}, 895-898 (1993).

\end{thebibliography}
\end{document}